\documentclass[aps,twocolumn,amsmath,amssymb]{revtex4}
\usepackage{graphicx}
\usepackage{color}
\usepackage{amsfonts}
\newcommand{\bear}{\begin{array}}  \newcommand{\eear}{\end{array}}
\newcommand{\bea}{\begin{eqnarray}}  \newcommand{\eea}{\end{eqnarray}}
\newcommand{\beq}{\begin{equation}}  \newcommand{\eeq}{\end{equation}}
\newcommand{\bef}{\begin{figure}}  \newcommand{\eef}{\end{figure}}
\newcommand{\bec}{\begin{center}}  \newcommand{\eec}{\end{center}}

\newcommand{\Eqn}[1]{&\hspace{-0.2em}#1\hspace{-0.2em}&}

\begin{document}
\title{  Comment on
``Einstein's Other Gravity and the Acceleration of the Universe''}

\author{Kazuharu Bamba,
Chao-Qiang Geng,
and 
Chung-Chi Lee
}
\affiliation{
Department of Physics, National Tsing Hua University, Hsinchu, Taiwan 300 
}

\begin{abstract}
We show that in the exponential $f(T)$ model studied by
E. Linder [Phys.\ Rev.\  D {\bf 81}, 127301 (2010)],
it is impossible to have
the crossing of the phantom divide 
line $w_{\mathrm{DE}}=-1$.
\end{abstract}

\pacs{
 98.80.-k, 04.50.Kd}

\maketitle

\begin{figure}[htbp]
\includegraphics*[width=2.5 in]{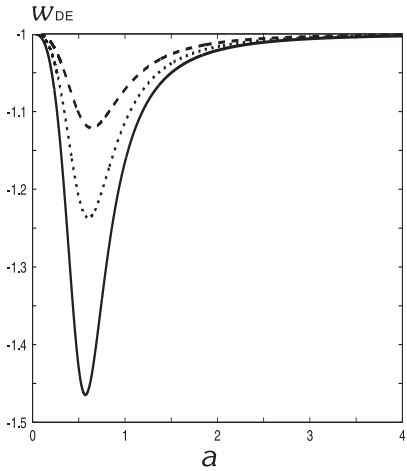}
\caption{Effective dark energy equation of state $w_{\mathrm{DE}}(a)$ as a function of the scale factor $a$ for the exponential $f(T)$ model,
where the dash, dotted and solid lines represent  $p=0.25$, 0.5 and 1, respectively.}
 \label{fig01}
\end{figure}

Recently, a new modified gravity theory, called as $f(T)$ with a torsion 
scalar $T$,
 has been proposed~\cite{Bengochea:2008gz,Linder:2010py}
to explain the present accelerating expansion of the universe. The new theory
is extended by the teleparallel gravity with the Weitzenb¬ock connection instead of
the Levi-Civita connection in  general relativity. 
In this new theory, the gravity is no longer caused by curved spacetime but torsion
and 
moreover, 
the field equations  are only second order unlike the fourth order equations in
the f(R) theory. 
In Ref.~\cite{Linder:2010py}, 
Linder has studied the exponential $f(T)$ model, given by
\begin{equation}
f \equiv f(T)=-\alpha T \left(1-e^{pT_0/T}\right)\,,
\label{eq:3.5}
\end{equation}
where 
%
$\alpha = [1-\Omega_{\mathrm{m}}^{(0)}]/[1-\left(1-2p\right)e^p]$, 
%
 $T_0=T(z=0)$ is the current torsion, and $p$ is a constant with $p=0$ corresponding to
 $\Lambda$CDM.
In particular, he  
showed that
 the  effective dark energy  equation of state~\cite{Linder:2010py}
\begin{equation}
w_{\mathrm{DE}}
=
-\frac{f/T-f_T+2Tf_{TT}}{\left(1+f_T+2Tf_{TT}\right)\left(f/T-2f_T\right)}
\label{W}
\end{equation}
crosses  the phantom divide 
line $w_{\mathrm{DE}}=-1$ for $p\leq 1$ from $w_{\mathrm{DE}}>-1$ to $w_{\mathrm{DE}}<-1$
in the opposite manner from the viable $f(R)$ models~\cite{BGL}.
However, we find out that that the crossing of the phantom divide 
could not be realized in the exponential $f(T)$ model in Eq.~(\ref{eq:3.5}). With the same set of the parameters
as in Ref.~\cite{Linder:2010py}, in Fig.~\ref{fig01} 
we show
$w_{\mathrm{DE}}$ in Eq.~(\ref{eq:3.5}) 
as a function of the scale factor $a$ for $p=0.25$ (dash), 0.5 (dotted), and 1 (solid).
 It is clear that the equation of state is always less than -1 in contrast to 
the results of Fig. 1 in 
  Ref.~\cite{Linder:2010py}.
In fact, it can be shown~\cite{BGL-1} that the phase of the universe 
depends on the sign of the parameter $p$, i.e., 
for $p<0$ the universe always stays in the non-phantom phase, 
whereas for $p>0$ it always in the phantom phase. Our results also agree with
those in Ref.~\cite{Wu:2010mn}.

We now discuss the reason why the crossing of the phantom divide cannot occur
 in the exponential $f(T)$ theory in Eq.~(\ref{eq:3.5}) for $p\leq 1$. 
{}From Eq.~(\ref{eq:3.5}), we find 
\begin{eqnarray}
f_T \Eqn{=} 
\frac{d f(T)}{d T} =- \alpha \left( 1-e^{pT_0/T}+\frac{pT_0}{T} 
e^{pT_0/T}\right)\,,
\label{eq:3.7} \\ 
f_{TT} \Eqn{=} \frac{d^2 f(T)}{d T^2} = \alpha \left(\frac{pT_0}{T}\right)^2 
\frac{1}{T} e^{pT_0/T}\,.
\label{eq:3.8}
\end{eqnarray}

To illustrate our results, we only concentrate on the limits of $0< p \ll 1$ and $X \equiv pT_0/T \ll 1$. 
In this case, $T_0/T \lesssim 1$, which corresponds to the region 
from the far past to the near future. 
Consequently, Eqs.~(\ref{eq:3.5}), (\ref{eq:3.7}) and (\ref{eq:3.8}) are approximately 
expressed as  
\begin{eqnarray}
\frac{f}{T} \approx \alpha \left(X+ \frac{X^2}{2}\right), 
f_T \approx - \frac{\alpha X^2}{2},
Tf_{TT} \approx \alpha X^2\,.
\label{eq:3.11}
\end{eqnarray}
%
Substituting Eq.~(\ref{eq:3.11}) into 
Eq.~(\ref{W}), we obtain 
\begin{eqnarray}
w_{\mathrm{DE}} 
\Eqn{\approx}  
-1-\frac{3X^2/2}{X+3X^2/2}\,,
\label{eq:3.13} 
\end{eqnarray}
where in deriving the approximate equality in Eq.~(\ref{eq:3.13}) 
we have used $\alpha \sim O(1)$. 
 From Eq.~(\ref{eq:3.13}) we see that 
the universe always stays in the phantom phase 
(because $p>0$ and $w_{\mathrm{DE}} < -1$). 

Finally, it is interesting to note that 
in the various $f(T)$ models extended from the viable $f(R)$ models in Ref.~\cite{Yang:2010hw}, 
all evolutions of $w_{\mathrm{DE}}$ show no crossing  the phantom divide.\\

\noindent
{\bf Acknowledgments}

One of the authors (CQG) would like to thank
the Institute of Theoretical Physics, Beijing  for hospitality and a wonderful 
program on ``Dark Energy and Dark Matter'' at Weihai,.
 The work is supported in part by 
the National Science Council of R.O.C. under
Grant \#:
NSC-98-2112-M-007-008-MY3
and 
NTHU under the Boost Program \#: 
99N2539E1.


\end{document}